# Negative Goos-Hänchen Shifts with Nano-metal-films on Prism Surface

Jiang Rong[1)]    Huang Zhi-Xun[1)]    Lu Gui-Zhen[1)]

1) Communication University of China, Beijing 100024, P. R. China

Corresponding authors. Email: tonytonychopper@cuc.edu.cn(Rong Jiang)

Abstract: In recent years, the fantastic phenomena of negative Goos-Hänchen shifts were studied in several optics experiments with the positive refractivity materials, which is predicted by many scientists through both classical physics or quantum physics theory. In order to verify the negative Goos-Hänchen shifts phenomena, an experiment in microwave frequency was done with nano-metal-films on prism surface. Because using nano-metal-films enhances self-interference effect, not only were the negative Goos-Hänchen shifts obtained, but also the giant Goos-Hänchen shifts was appeared.

1.    Introduction

As well known, when a beam of electromagnetic wave is obliquely incident upon the single interface between an optically denser medium, the Goos-Hänchen(GH) shifts[1], will be produced due to some energy-flow transfers from higher reflectivity medium to lower reflectivity medium, and eventually comes back into the higher reflectivity medium[2]. The special phenomenon has been extensively researched in the past 50-old years, whatever theoretical or experimental, great progress has been made by lots of academic team[3-6].Especially, people found that the penetrating Poynting-energy sometimes forward propagate along the interface, but sometimes amazingly backward along the interface. The forward energy naturally produce forward surface waves and the positive GH shifts, and backward one correspondingly produce backward moving surface waves and the negative GH shifts[7], as shown in Fig.1. However, the fantastic phenomena of negative GH shifts just be possible to interpret the essence of shifts effect at an interface. The theory of negative GH shifts was predicted by T. Tamir[8], and in 1978.B.A.Anicin[9] computed the negative GH shifts of corrugated metallic surface by application of the stationary phase method. Hereafter, several theoretical calculations have since been done to explain the phenomenon because of the development of the surface wave theory. For example, W. Wild[10]  in1982,   and

J.Zhang[11] in 2008, calculated and derived negative GH shifts of smooth metallic surface in the microwave and visible portions of the spectrum. There are many studies of negative GH shifts, but the analytical method is mainly primeval stationary phase approach by K. Artmann[12] and energy flux method by R. H. Renard[2]. To remedy this, Yasumoto and Oishi[13] proposed a "new" energy flux method, and Bai and Zhang[14] proposed a new method studying wedge-shaped.

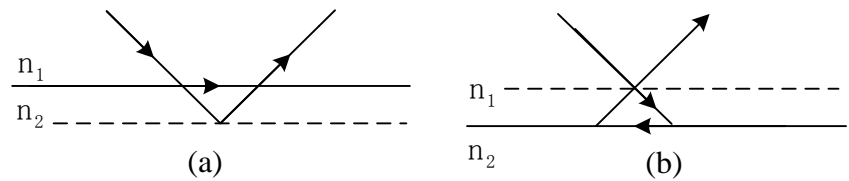

Fig.1 This is the GH shifts in two cases. (a) The Poynting-energy forward propagate along the interface. The positive GH shifts was formed (b) The Poynting-energy forward propagate along the interface. The negative GH shifts was formed.

As far as we know, the first experiment of negative GH shifts was observed on a liquid-solid interface with superimposed grating at 6MHz ultrasonic beam by M.Breazeale[15] in 1976. But this experiment has not been extensive concerned. In recent years, with the development of the technology the research of negative GH shifts emerged at a new climax. The phenomenon of negative GH shifts was observed on periodic structures interface[16], on metal films interface[17,18], on negative refraction materials[19], on weakly absorbing dielectric film[20], and on thin-film[21]. These experiments demonstrated that in the case of TM polarized light beam it is not difficult to obtain negative and giant GH shifts. In 2008, R. Gruschinski[22] first used metal films to do the experiment in the case of TE polarized microwaves. However, they only observed giant GH shifts, but without negative GH shifts. And in 2009 Zhang, Wen and Zhang[23] use a novel method that is used two prism obtain giant GH shifts in total internal reflection in TE polarized, too. In 2010, Qu and Huang[24] used EBG(Electromagnetic Band Gap)structure obtain negative GH shifts in TE polarized microwave beam. Based on these studies, in this paper we focus mainly on the GH shifts experiment at nano-metal-films surface in microwaves frequency.

Experimental system used Kretschmann's structure[25]. The aluminum films with thickness 30nm and 60nm on prism surface were studied. This system is used to observe negative GH shifts on nano-metal-films surface in microwave frequency. The GH shifts is investigated for various frequencies, various thickness of metal films, various incident angles and different polarizations. In our experiment, the negative GH shifts and giant GH shifts was observed.

2. Experimental arrangement

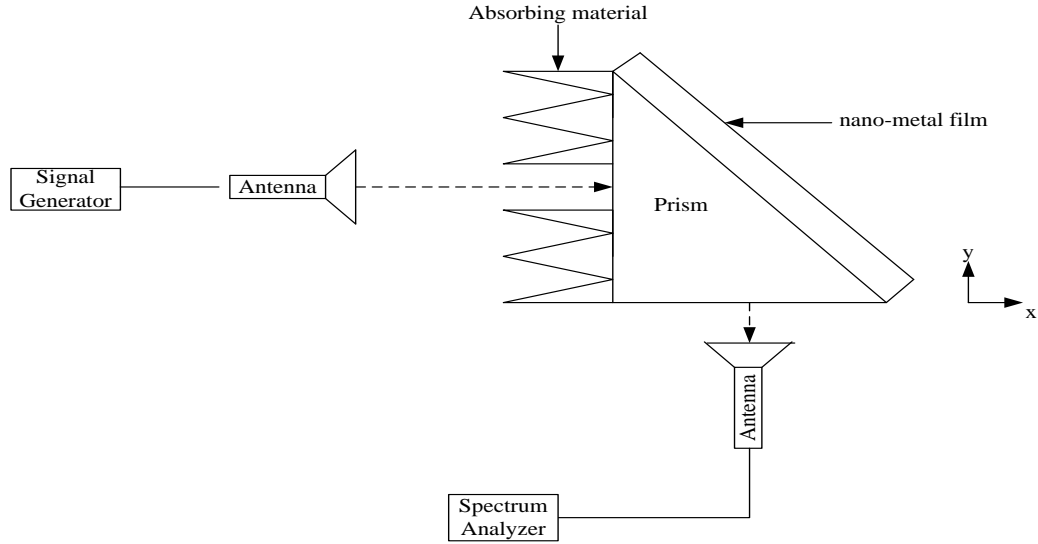

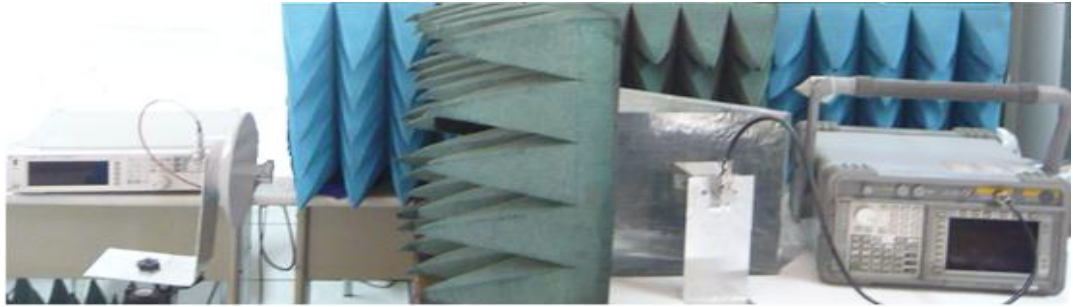

Fig. 2. Schematic diagram of experimental set-up

A diagram of experimental system is illustrated in Fig. 2 (with the photo). The microwave signal is generated by Agilent MXG-N5183A Signal Generator, and fed into a linear polarization parabolic antenna whose diameter is 20cm, and the operation frequency is (8.2-12.4)GHz. To obtained bounded collimation beam, the distance between antenna and prism is more than 150cm, and absorbing material with a rectangle aperture of $10cm \times 20cm$ is added in front of prism(the size of aperture is determined by several experiment). The prism is made by PMMA with the dimension of $32cm \times 32cm \times 30cm$, and the refractive index is approximately 1.6 corresponding to a critical angle of $38.7°$. The nano-metal-films are closely placed on the reflecting surface, that is the oblique planes of prism. The nano-metal-films are produced by aluminum evaporated on 18μm polyethylene substrates. The layer thickness, including the substrate of all the applied films, is much smaller than the wavelength. The polyethylene substrate has little influence on experiment. The reflection wave is received by the horn of 6cm × 8.4cm with the same polarization and frequency as the incident antenna. And the horn is connected to Agilent E4407b Spectrum Analyzer.

3. Experimental results

First, we used the Al-film with 30nm in the experiment. The angle of incidence is $45°$, and it is greater than critical angle. The GH shift was measured as frequency changing in the range of 8.2-12.4GHz at TE polarization. The results are shown in Fig. 3. It is indicated that the GH shifts are all the negative shifts.

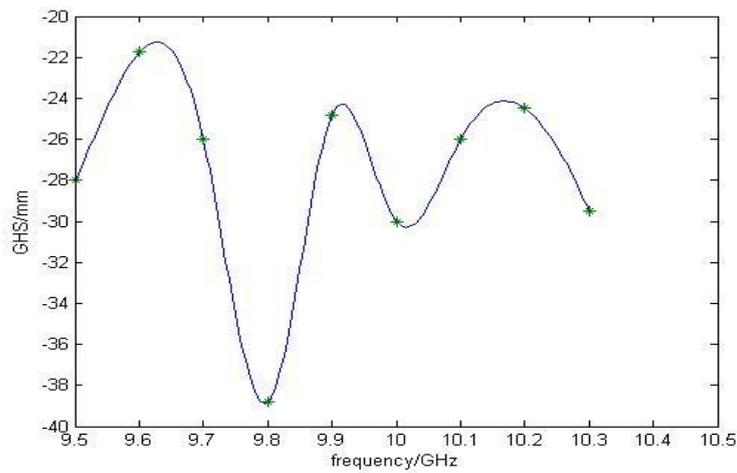

Fig. 3. Experimental data of GH shifts is on 30nm film surface in TE polarization

Then, we altered the polarization to measure the shift, but the negative shift was not obtained at TM polarization, as shown in the Fig. 4.

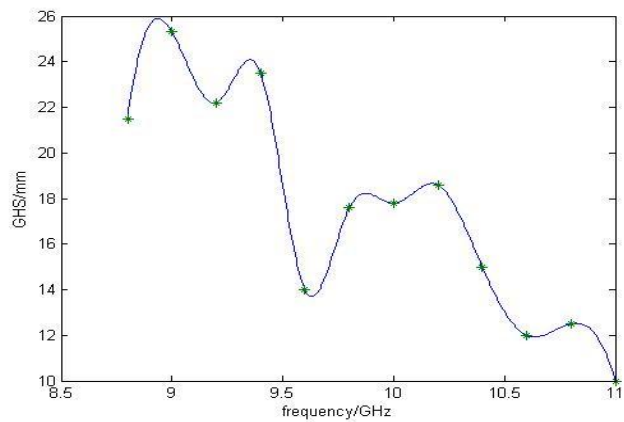

Fig. 4. Experimental data of GH shifts is on 30nm film surface at TM polarization

These peculiar results are due to the special character of nano-metal-film mentioned above. Then these results contrasted with the results without nano-metal-films in our previous experiment that show in Fig. 5 and Fig. 6.

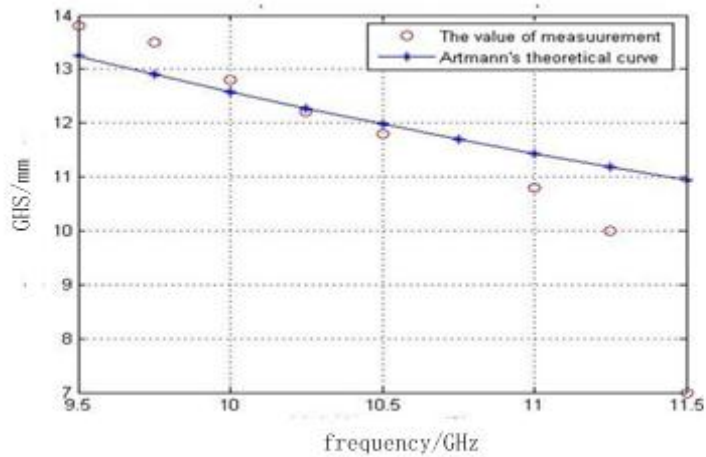

Fig. 5. Experimental data of GH shifts in TE polarization without nano-metal-films.

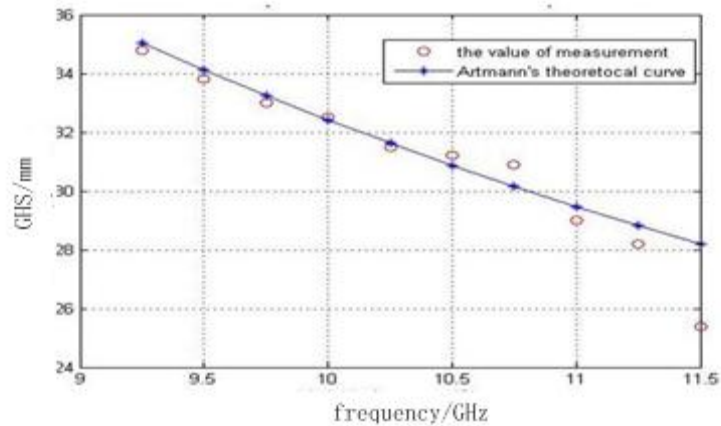

Fig. 6. Experimental data of GH shifts in TM polarization without nano-metal-films.

Through the above comparison, we can see that when nano-metal-films were used in the experiment, negative GH shifts were observed at TE polarization. This phenomenon could occur because using nano-metal-films make self-interference shift enhancement. we all know that in Yasumoto and Oishi's[13] energy flux model the self-interference shift differ by GH shifts given by Artman's and Renard's, which is due to the interaction between the incident and reflected beams. And in some special construction this self-interference effect can be enhanced[21]. Because the microstructure and properties of nano-metal-film(less than 100nm) is different from the microscopic system of atoms and molecules, and is also different from the macroscopic system shown the intrinsic characteristics of lager granular materials. The core part of it is small an can't be regarded as an infinite long-range order structure, while the number of surface atoms increase, and is even more than the number of bulk phase atoms, it makes the particles surface activity increase. Therefore, due to the unusual structure surface-interface effect, small size effect, macroscopic quantum phenomena are generated. Because of these peculiar properties make the self-interference effect enhancement. We know that self-interference enhancement make reflection weak. So the reflection coefficient was measured shown in the fig. 7.

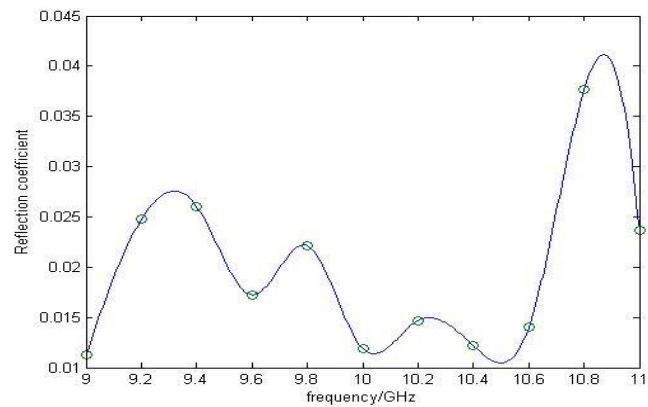

Fig. 7. Reflection coefficientis on 30nm film surface at TE polarization

To better understand the impact of the special nano-metal-films on GH shifts, the thickness of metal film was increased(but the film is very thin too, less than 100nm). The Al-film with 60nm was used in experiment, and the thickness of the film is much smaller than the wavelength. The same method was used to place the Al-film with 60nm. Meanwhile, Experiment was carried out at the same frequency band in TE and TM polarizations. The results are shown in Fig. 8 and Fig. 9.

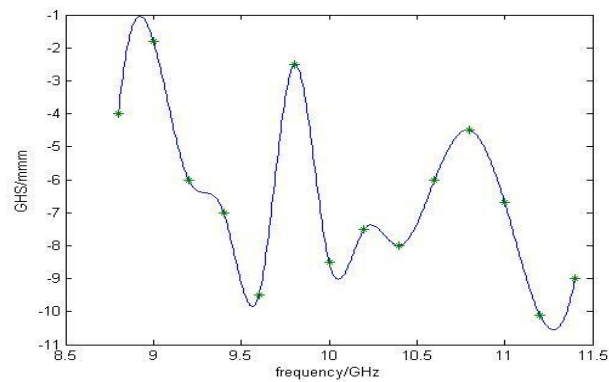

Fig. 8. Experimental data of GH shifts is on 60nm film surface in TE polarization

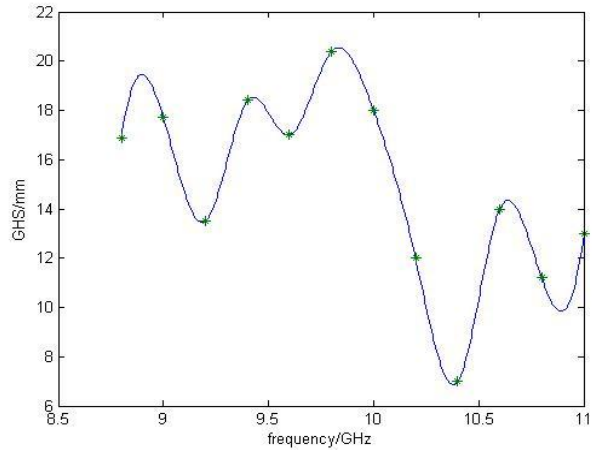

Fig. 9. Experimental data of GH shifts is on 60nm film surface at TM polarization

Because the film is enough thin, the GH shifts is also negative. But the absolute value of the GH shift is reduced with thickness of film increased, and when the film is very thick(more than 1000nm), we almost can't measure the GH shift(i.e. GH shift=0).

As we know, the angle of incident influences the general GH shift. So in order to investigate better the impact of nano-metal-films on GH shifts, GH shifts was measured in various different incident angles. The 9GHz frequency is used in the experiment. The results are shown in Fig. 10.

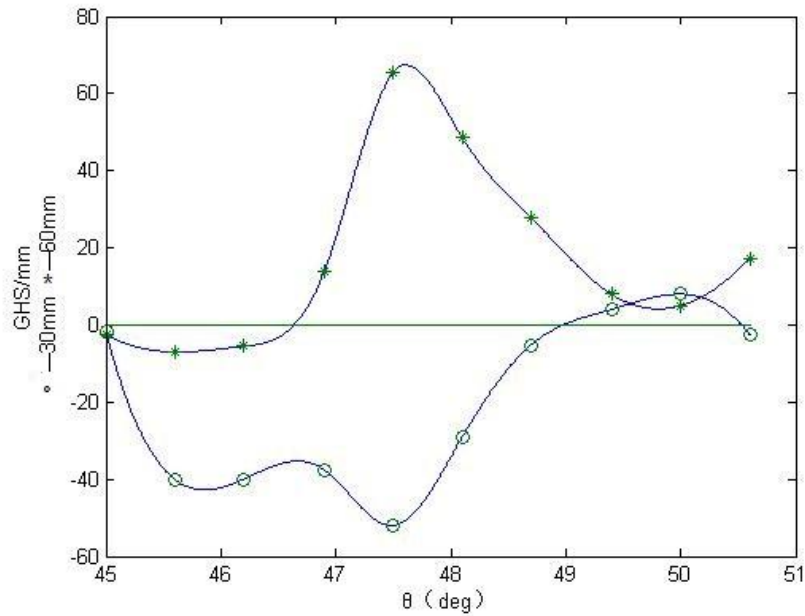

Fig. 10. Experimental data of GH shifts is on 30nm and 60nm films surface in TE polarization in various different incident angles.

In Fig.10 we can see that both positive GH shifts and negative GH shifts were observed in different angles on the same metal film surface. When the thickness of the metal film is 30nm, GH shifts is mainly negative. When the thickness of the metal film is 60nm, GH shifts is mainly positive. Only at incident angle less than 46.5°, GH shifts is negative. When the incident angle was fixed at 45°, GH shifts is negative used two kinds of films. However, when the incident angle was more than 45°, GH shifts may be positive. And when the incident angle was approximate 47.5°, both the absolute value of negative GH shifts with 30nm film and the value of the positive GH shifts with 60nm film are maximum value, that is the giant shifts said by I.Shadrivov[26]. Because of changing incident angle influence make the times of reflection increased[14], the self-interference[27] was enhanced. So the GH shift on the surface of nano-metal-films is larger in some incident angle, and with the increase of film's thickness the special character of the film is unobvious, so the absolute value of negative GH shifts is smaller.

Summary

Because nano-metal-films have unusual electromagnetic character, including surface-interface effect, small size effect, macroscopic quantum phenomena, the inhomogeneity of nano-metal-films thickness, etc. The negative GH shifts was observed on the surface of nano-metal-films with the thickness less than wavelength. Negative GH shifts and positive GH shifts were all obtained by changing the incident angle, and the giant GH shifts was obtained at incident angle of 47.5°.

This word was supported by Specialized Research Fund for the Doctoral Program of Higher Education (No. 200800330002)